\documentstyle[12pt,twoside,fleqn,espcrc1,epsfig]{article}

\newcommand{\AmS}{{\protect\the\textfont2
  A\kern-.1667em\lower.5ex\hbox{M}\kern-.125emS}}

\title{Light and strange baryons, two-baryon systems 
and the chiral symmetry of QCD}

\author{L. Ya. Glozman\address{Institute for Theoretical Physics, 
        University of Graz, A-8010 Graz, 
Austria}\address{Department of Physics, University of Helsinki, POB9,
FIN-00014 Helsinki, Finland}}%

\begin{document}
\maketitle

\begin{abstract}
Beyond the scale of spontaneous breaking of chiral symmetry 
light and strange baryons should be considered
as systems of three constituent quarks with confining interaction and
a chiral interaction that is mediated by Goldstone bosons between the
constituent quarks. The flavor-spin structure and sign of
the short-range part of the Goldstone boson exchange interaction reduces the
$SU(6)_{FS}$ symmetry down to $SU(3)_F \times SU(2)_S$, induces hyperfine
splittings and provides correct ordering of the lowest states with positive
and negative parity. A unified description of 
light and strange baryon spectra calculated in a semirelativistic framework
is presented.  It is demonstrated that the same 
short-range part of the Goldstone boson exchange between 
the constituent quarks 
 induces a strong short-range repulsion in $NN$ system when the
latter is treated as $6Q$ system.
Similar to the $NN$ system there should be a short-range 
repulsion in other $NY$ and $YY$ two-baryon systems. 
We also find that the compact 6Q system
with the "H-particle" quantum numbers lies a few hundreds MeV above the
$\Lambda\Lambda$ threshold. It then suggests that the deeply bound H-particle
should not exist. 
\end{abstract}

\section{Spontaneous Breaking of Chiral Symmetry  and its Implications}

 At low temperatures and densities the  
$SU(3)_{\rm L} \times SU(3)_{\rm R}$ chiral symmetry of QCD Lagrangian
is spontaneously
broken down to $SU(3)_{\rm V}$ by the QCD vacuum (in the large $N_c$ limit
it would be $U(3)_{\rm L} \times U(3)_{\rm R} \rightarrow U(3)_{\rm V}$).
 A direct evidence for the spontaneously broken
chiral symmetry is a nonzero value of the quark condensates for the
light flavors
$<|\bar{q}q|> \approx -(240-250 {\rm
MeV})^3,$
which represent the order parameter. That this is indeed so, we know from
three  sources: current algebra,
QCD sum rules, and lattice gauge calculations.
There are two important generic consequences of the spontaneous breaking
of chiral symmetry (SBCS). The first one is an appearance of the octet
of pseudoscalar mesons of low mass, $\pi, {\rm K}, \eta$, which represent
the associated approximate Goldstone bosons (in the large $N_c$ limit the
flavor singlet state $\eta'$ should be added). The second one is that valence
(practically massless) quarks acquire a dynamical mass, which have been called 
historically  constituent quarks. Indeed, 
the nonzero value of the quark condensate 
itself implies at the formal level that there should be rather big
dynamical mass, which could be in general a moment-dependent quantity 
\cite{Mir}. Thus the constituent quarks should be considered as quasiparticles
whose dynamical mass comes from the nonperturbative 
gluon and quark-antiquark dressing.
The flavor-octet axial current conservation in the chiral limit tells that
the constituent quarks and Goldstone bosons should be coupled with the
strength $g=g_A M/f_\pi$ \cite{WEIN}, which is a quark analog of the famous 
Goldberger-Treiman relation. We cannot say at the moment for sure what is the
microscopical mechanism for SBCS in QCD. Any sufficiently strong
scalar interaction between quarks will induce the SBCS (e.g. the 
instanton - induced interaction contains the scalar part, or it can be
generated by monopole condensation, etc.).
The most general aspects of SBCS 
such as constituent quark mass generation, 
Goldstone boson creation and their couplings are well illustrated
by the Nambu and Jona-Lasinio model \cite{Nambu}.
For the low-energy baryon properties it is only essential that
beyond the spontaneous chiral symmetry breaking scale (i.e. at low resolution)
 new
dynamical degrees of freedom appear - constituent quarks and chiral
fields which couple together. The low-energy baryon 
properties are mainly determined
by these dynamical degrees of freedom and the confining interaction.

We have recently suggested \cite{GLO2} that in the low-energy regime 
light and strange baryons
should be considered as a system of 3 constituent quarks with an effective
$Q-Q$ interaction that is formed of central confining part and a chiral
interaction mediated by the Goldstone bosons between constituent quarks.
This physical picture allows to understand a structure of the baryon spectrum
and to solve, in particular,  the long-standing problem of ordering of the
lowest baryons with positive- and negative-parity.

\section{The Goldstone Boson Exchange Interaction 
and the Structure of  Light and Strange Baryon Spectra}

The
coupling of the constituent  quarks and the pseudoscalar Goldstone
bosons will (in the $SU(3)_{\rm F}$ symmetric approximation) have
the form $g/(2m)\bar\psi\gamma_\mu
\gamma_5\vec\lambda^{\rm F}
\cdot \psi \partial^\mu\vec\phi$ within the nonlinear realization of chiral
symmetry (it would be $ig\bar\psi\gamma_5\vec\lambda^{\rm F}
\cdot \vec\phi\psi$ within the linear $\sigma$-model chiral symmetry 
representation). A coupling of this
form, in a nonrelativistic reduction for the constituent quark spinors,
will -- to lowest order -- give rise the 
$\sim\vec\sigma \cdot \vec q \vec\lambda^{\rm F}$ structure of the 
meson-quark vertex, where $\vec q$ is meson momentum. Thus, the structure
of the  potential between quarks "$i$" and "$j$" in momentum representation is

\begin{equation}V(\vec q) \sim
\vec\sigma_i\cdot\vec q \sigma_j\cdot\vec q ~
\vec\lambda_i^{\rm F}\cdot\vec\lambda_j^{\rm F}
D(q^2) F^2(q^2)
,\label{2} \end{equation}

\noindent
where $D(q^2)$ is dressed Green function for chiral field which includes
both nonlinear terms of chiral Lagrangian and fermion loops, $F(q^2)$ is
meson-quark formfactor which takes into account the internal structure
of quasiparticles. At big distances ($\vec q \rightarrow 0$), one has 
$D(q^2) \rightarrow D_0({\vec q}^2)= -({\vec q}^2 + \mu^2)^{-1}
\not= \infty $ and $F(q^2) \rightarrow 1$.
 It then follows from
(\ref{2}) that $V(\vec q = 0) = 0$, which means that the volume integral
of the Goldstone boson exchange (GBE) interaction should vanish, 

\begin{equation}
\int d\vec r V(\vec r) = 0. 
\label{1} \end{equation}

\noindent
This sum rule is not valid in the chiral limit,
however, where $\mu =0$ and, hence, $D_0(\vec q =0)=\infty$.
Since at big interquark separations the spin-spin component of the
pseudoscalar-exchange interaction is 
$V(r) \sim {e^{-\mu r}\over{r}}$, 
it then follows from the sum rule above
 that at short interquark separations the spin-spin
interaction should be opposite in sign as compared to the Yukawa tail and
very strong. {\it It is this short-range part of the Goldstone boson exchange
(GBE) interaction between the constituent quarks that is of crucial importance
for baryons: it has a sign appropriate to reproduce the level splittings
and dominates over the Yukawa tail towards short distances.} In a
oversimplified consideration with a free Klein-Gordon Green function instead
of the dressed one in (\ref{2}) and with $F(q^2)=1$, one obtains the
following  spin-spin component of $Q-Q$ interaction:

\begin{equation}V(r)=
\frac{g^2}{4\pi}\frac{1}{3}\frac{1}{4m_im_j}
\vec\sigma_i\cdot\vec\sigma_j\vec\lambda_i^{\rm F}\cdot\vec\lambda_j^{\rm F}
\{\mu^2\frac{e^{-\mu r}}{ r}-4\pi\delta (\vec r)\}
.\label{3} \end{equation}

\noindent
In the chiral limit only the negative short-range part of the GBE interaction
survives.

 Consider first, for the purposes of illustration, a schematic model
which neglects the radial dependence
of the potential function $V(r)$ in (\ref{3}), and assume a harmonic
confinement among quarks as well as $m_{\rm u}=m_{\rm d}=m_{\rm s}$.
In this model

\begin{equation}H_\chi = -\sum_{i<j}C_\chi~
\vec \lambda^{\rm F}_i \cdot \vec \lambda^{\rm F}_j\,
\vec
\sigma_i \cdot \vec \sigma_j.\label{4} \end{equation}

\noindent
Note, that contrary to the color-magnetic interaction from perturbative
one-gluon exchange, the GBE interaction is explicitly flavor-dependent.
It is this circumstance which allows to solve the long-standing problem
of ordering of the lowest positive-negative parity states.

If the only interaction between the
quarks were the flavor- and spin-independent harmonic confining
interaction, the baryon spectrum would be organized in multiplets
of the symmetry group $SU(6)_{\rm FS} \times U(6)_{\rm conf}$. In this case
the baryon masses would be determined solely by the orbital structure,
and the spectrum would be organized in an {\it alternative sequence
of positive and negative parity states,} i.e. in this case the spectrum
would be: ground state of positive parity 
($N=0$ shell, $N$ is the number of harmonic 
oscillator excitations in a 3-quark state), first excited band of
negative parity ($N=1$), second excited band of positive parity ($N=2$), etc.

The Hamiltonian (\ref{4}), within a first order perturbation theory,
 reduces the $SU(6)_{\rm FS} \times U(6)_{\rm conf}$ symmetry down to
 $SU(3)_{\rm F}\times SU(2)_{\rm S}\times U(6)_{\rm conf}$, which automatically
implies a splitting between the octet and decuplet baryons (e.g. the $\Delta$
resonance becomes heavier than nucleon).

Let us now see how the pure confinement spectrum above becomes modified when
the GBE Hamiltonian (\ref{4}) is switched on.
For the octet states ${\rm N}$, $\Lambda$, $\Sigma$,
$\Xi$ ($N=0$ shell) as well as for their first
radial excitations of positive parity
 ${\rm N}(1440)$, $\Lambda(1600)$, $\Sigma(1660)$,
$\Xi(?)$ ($N=2$ shell) 
 the expectation value of the
Hamiltonian (\ref{4})
 is $-14C_\chi$. For the decuplet states
$\Delta$, $\Sigma(1385)$, $\Xi(1530)$, $\Omega$ ($N=0$ shell)
the corresponding matrix element is
$-4C_\chi$. In the  negative parity excitations
($N=1$ shell) in the ${\rm N}$, $\Lambda$ and $\Sigma$ spectra 
(${\rm N}(1535)$ - ${\rm N}(1520)$, $\Lambda(1670)$ - $\Lambda(1690)$
and $\Sigma(1750)$ - $\Sigma(?)$)
the contribution of the interaction (\ref{4})  is $-2C_\chi$. 
The first negative
parity excitation in the $\Lambda$ spectrum ($N=1$ shell)
$\Lambda(1405)$ - $\Lambda(1520)$ is flavor singlet 
and, in this case, the corresponding matrix element is $-8C_\chi$. The latter
state is unique and is absent in other spectra due to its flavor-singlet
nature.

These  matrix elements alone suffice to prove that
the ordering of the lowest positive and negative parity states
in the baryon spectrum will be correctly predicted by
the chiral boson exchange interaction (\ref{4}).
The constant $C_\chi$ may be determined from the
N$-\Delta$ splitting to be 29.3 MeV.
The oscillator
parameter $\hbar\omega$, which characterizes the
effective confining interaction,
may be determined as  one half of the mass differences between the
first excited
$\frac{1}{2}^+$ states and the ground states of the baryons,
which have the same flavor-spin, flavor and spin symmetries
(e.g. ${\rm N}(1440)$ - ${\rm N}$, $\Lambda(1600)$ - $\Lambda$, $\Sigma(1660)$
- $\Sigma$),
to be
$\hbar\omega \simeq 250$ MeV. Thus the two free parameters of this simple model
are fixed and we can  make now predictions.

In the ${\rm N}$, $\Lambda$  and $\Sigma$ sectors the mass
difference between the lowest
excited ${1\over 2}^+$ states (${\rm N}(1440)$, $\Lambda(1600)$, 
and $\Sigma(1660)$)
and ${1\over 2}^--{3\over 2}^-$ negative parity pairs
 (${\rm N}(1535)$ - ${\rm N}(1520)$, $\Lambda(1670)$ - $\Lambda(1690)$,
 and $\Sigma(1750)$ - $\Sigma(?)$, respectively) will then
be
\begin{equation}{\rm N},\Lambda,\Sigma:
\quad m({1\over 2}^+)-m({1\over 2}^--{3\over
2}^-)=250\, {\rm
MeV}-C_\chi(14-2)=-102\, {\rm MeV},\end{equation}
whereas for the lowest states in the $\Lambda$ system ($\Lambda(1600)$,
$\Lambda(1405)$ - $\Lambda(1520)$) it should be

\begin{equation}\Lambda:\quad m({1\over 2}^+)-m({1\over 2}^--{3\over
2}^-)=250\, {\rm
MeV}-C_\chi(14-8)=74\, {\rm MeV}.  \end{equation}

This simple example shows how the GBE interaction 
provides different ordering of the lowest positive and negative parity excited
states in the spectra of the nucleon and
the $\Lambda$-hyperon. This is a direct
consequence of the symmetry properties of the boson-exchange interaction
\cite{GLO2}.
Namely, completely symmetric FS state in the ${\rm N}(1440)$,
$\Lambda(1600)$ and
$\Sigma(1660)$ positive parity resonances from the $N=2$ band feels a
much stronger
attractive interaction than the mixed symmetry state  in the
${\rm N}(1535)$ - ${\rm N}(1520)$, $\Lambda(1670)$ - $\Lambda(1690)$
and $\Sigma(1750)$ -$\Sigma(?)$ resonances of negative parity ($N=1$ shell).
Consequently the masses of the
positive parity states ${\rm N}(1440)$, $\Lambda(1600)$  and
$\Sigma(1660)$ are shifted
down relative to the other ones, which explains the reversal of
the otherwise expected "normal ordering".
The situation is different for $\Lambda(1405)$ - $\Lambda(1520)$
and
$\Lambda(1600)$, as the flavor state of  $\Lambda(1405)$ - $\Lambda(1520)$ is
totally antisymmetric. Because of this the
$\Lambda(1405)$ - $\Lambda(1520)$ gains an
attractive energy, which is
comparable to that of the $\Lambda(1600)$, and thus the ordering
suggested by the confining oscillator interaction is maintained.

Note that the problem of the relative position of positive-negative
parity states cannot be solved with other types of hyperfine interactions
(the colour-magnetic and instanton-induced ones).

\section{Semirelativistic Chiral Constituent Quark Model}

In the semirelativistic chiral constituent quark model \cite{GPVW}
the dynamical part of the Hamiltonian consists of linear pairwise
confining interaction and the GBE interaction, which includes both
the short-range part,parametrized phenomenologically, of the
form $ \sim -\Lambda^2 {e^{-\Lambda r}\over{r}}$, and the long-range
Yukawa tail. Both the flavor-octet ($\pi, K, \eta$) and flavor-singlet
($\eta'$) meson exchanges are taken into account. The coupling constant
of the flavor-octet mesons to constituent quarks is fixed from the known 
pion-nucleon coupling constant, while the $\eta'$-quark coupling constant
is treated phenomenologically.

The kinetic-energy operator is taken in relativistic form,
$H_0 = \sum_{i=1}^3 \sqrt(p_i^2 + m_i^2)$. The semirelativistic 
three-quark Hamiltonian was solved along the stochastical variational
method \cite{VARGA} in momentum space. For the whole Q-Q potential the
model involves a total of 4 free parameters whose numerical values are
determined from the fit to all 35 confirmed low-lying states. 
The string tension of the confining interaction
can be fixed to that one obtained from charmonium and bottomonium
spectroscopy and lattice calculations.

In fig. 1 we present the ground states as well as low-lying excited states
in $N$, $\Delta$, $\Lambda$, $\Sigma$, $\Xi$, and $\Omega$ spectra.
From the results of fig. 1 it becomes evident that within the chiral
constituent quark model a unified description of both nonstrange and
strange baryon spectra is achieved in good agreement with phenomenology.

It is instructive to learn how the GBE interaction affects the energy levels 
when it is switched on and its strength is gradually increased (Fig. 2).
Starting out from the case with confinement only, one observes that
the degeneracy of states is removed and the inversion of ordering
of positive and negative parity states 
is achieved in the $N$ spectrum, as well as for some states in the
$\Lambda$ spectrum, while the ordering of the lowest positive-negative
parity states is opposite in $N$ and $\Lambda$ spectra. The reason for
this behaviour is the flavor-spin structure and sign of the short-range
part of GBE.

\begin{figure}
\begin{center}
\epsfig{file=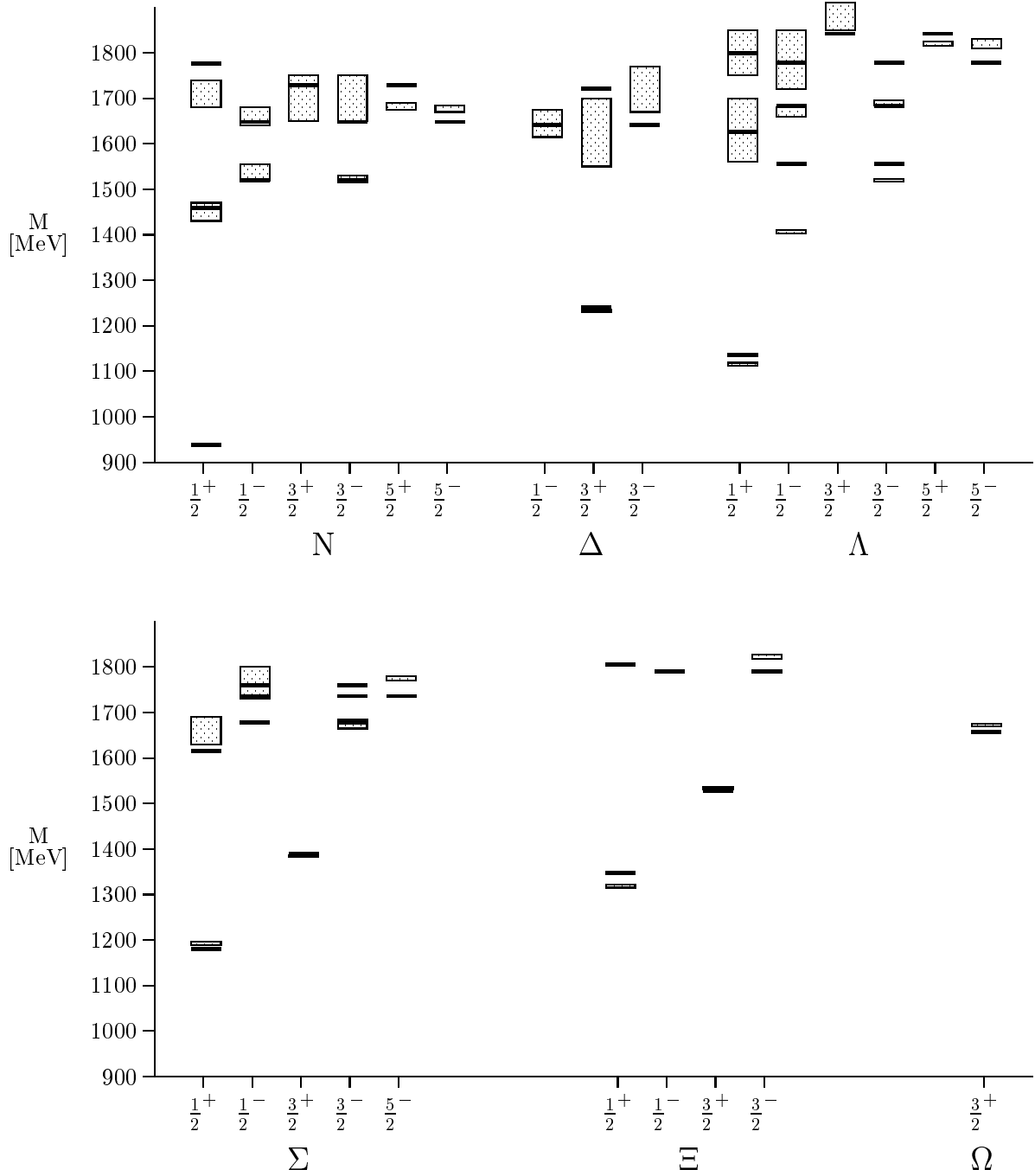}
\caption{Energy levels of the lowest light and strange baryon states 
(below 1850 MeV) with total
angular momentum and parity $J^P$. The shadowed boxes represent the experimental
values with their uncertainties.}
\end{center}
\end{figure}
\begin{figure}
\epsfig{file=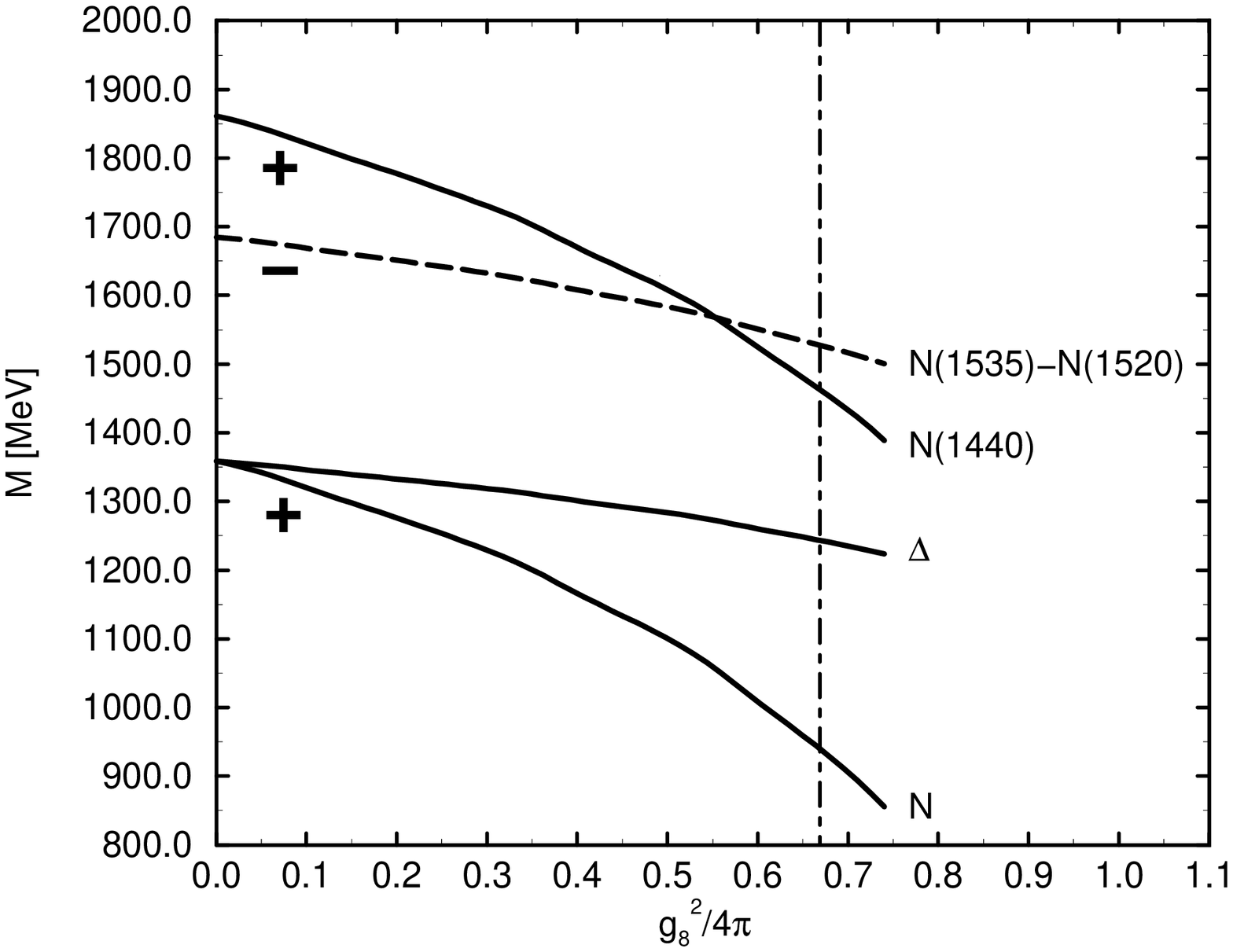,width=8cm}\hfill
\epsfig{file=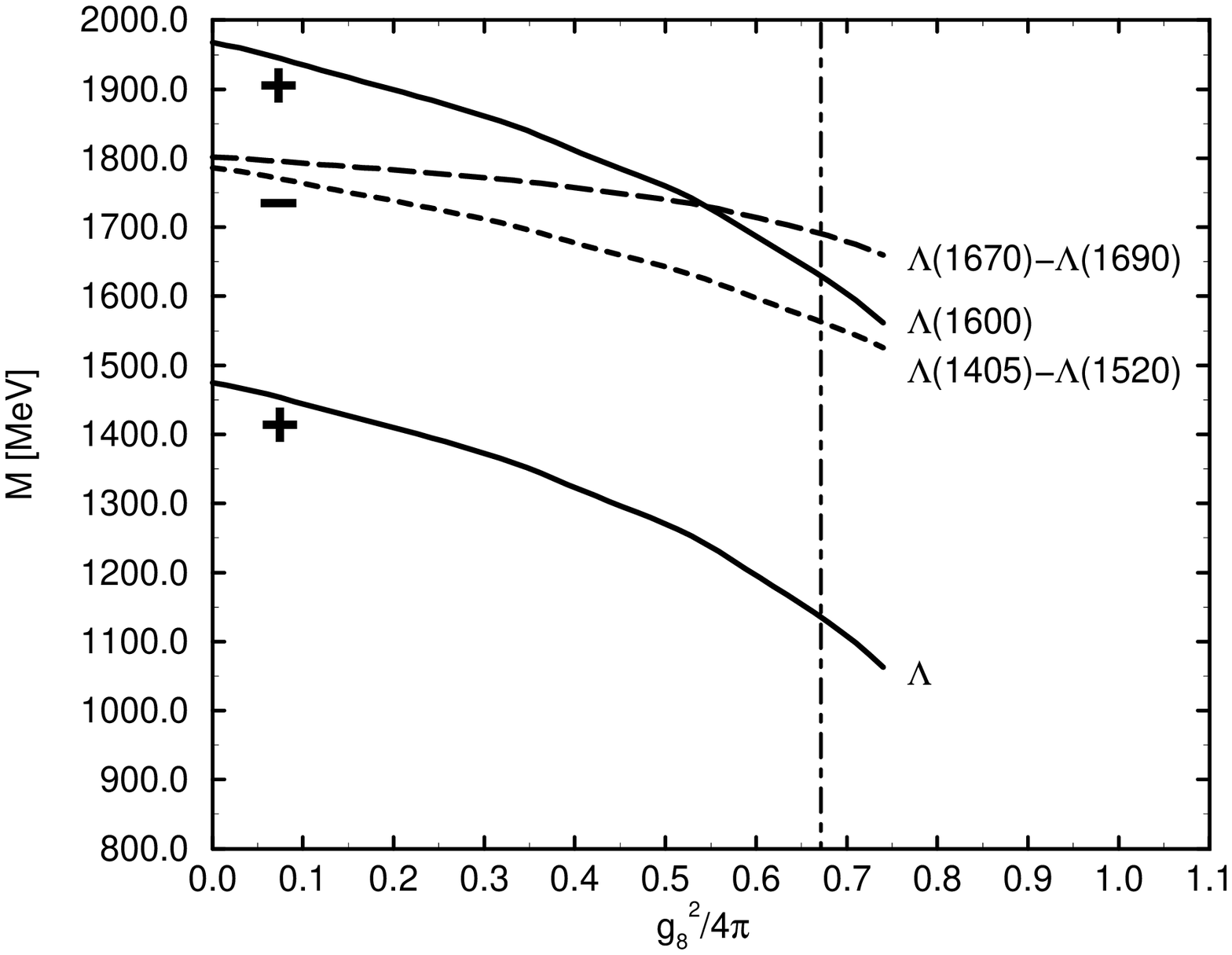,width=8cm}
\caption{Level shifts of some lowest baryons as a function of the 
strength of the GBE. Solid and dashed lines correspond to positive- 
and negative-parity states, respectively.}
\end{figure}
\begin{figure}
\begin{center}
\epsfig{file=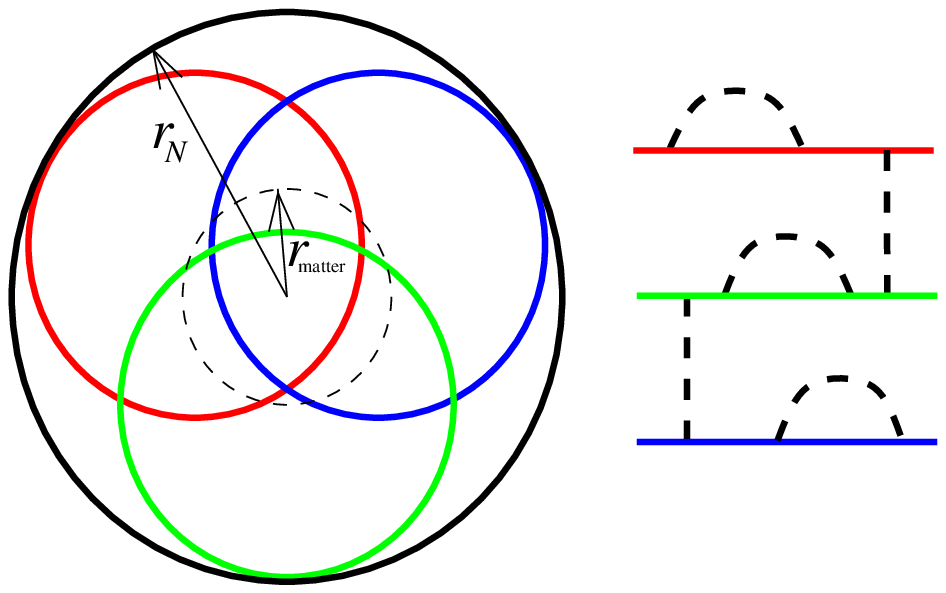}
\caption{Nucleon as it is seen in the low-energy and low-resolution regime.} 
\end{center}
\end{figure}

It is clear that the Fock components $QQQ\pi, QQQK, QQQ\eta$, and $QQQ\eta'$
(including meson continuum) cannot be completely integrated out in favour
of the meson-exchange $Q-Q$ potentials for some states above or near
the corresponding meson thresholds. Such  components in addition to the
main one $QQQ$ could explain e.g. an exceptionally big splitting of the
flavor singlet states $\Lambda(1405)-\Lambda(1520)$, since the $\Lambda(1405)$
lies below the $\bar K N$ threshold and can be presented as  $\bar K N$ bound
system \cite{DAL}. Note, that in the case of the present approach this old
idea is completely natural and does not contradict a flavor-singlet  $QQQ$ 
nature of $\Lambda(1405)$, while it would be in conflict with naive constituent
quark model where no room for mesons in baryons. 
An admixture of such components
will be  important in order to understand strong decays of some excited states.
While technically inclusion of such  components in addition to the main
one $QQQ$ in a coupled-channel approach is rather difficult task, it should
be considered as one of the most important future directions.

What is an intuitive picture of the nucleon in the low-energy regime? 
Due to some complicated nonperturbative gluodynamics quarks and antiquarks
in the nucleon sea are strongly correlated into virtual Goldstone bosons.
These Goldstone bosons in turn couple to valence quarks. Thus the
nucleon consists mostly of 3 constituent quarks which are very big objects due 
to their meson clouds (see fig. 3). These constituent quarks are all the time 
in strong overlap inside the nucleon. That is why the short-range part of 
GBE interaction 
(which is represented by the contact term in the oversimplified 
representation (\ref{3})) is so crucially important inside baryons.
When constituent quarks are well separated and there is a phase space for pion
propagation, the long-range Yukawa tail of GBE interaction as well as
the correlated two-pion exchanges become very important. It is these parts
of meson exchange which produce the necessary long- and intermediate-range 
attraction in two-nucleon system.

\section{The Baryon-Baryon Interaction in a Chiral Constituent Quark Model}

So far, all studies of the short-range $NN$ interaction within the
constituent quark model were based on the one-gluon exchange interaction
between quarks. They explained the short-range repulsion in the $NN$ system
as due to the colour-magnetic part of OGE combined with quark interchanges
between 3Q clusters \cite{OKA}. It has been shown, however, that there
is practically no room for colour-magnetic interaction in light 
baryon spectroscopy and any appreciable amount of colour-magnetic interaction, 
in addition to GBE, destroys the spectrum \cite{GPPVW}. This conclusion
is confirmed by recent lattice QCD calculations \cite{LIU}. If so, the
question arises which interquark interaction is responsible for the
short-range $NN$ repulsion. Below I show that the same short-range part
of GBE which causes e.g. $N-\Delta$ splitting and produces good baryon spectra,
also induces a short-range repulsion in $NN$ system when the latter is
treated as 6Q system \cite{STPEPGL}.

At present one can use only a simple nonrelativistic $s^3$ ansatz for
the nucleon wave function when one applies the quark model to $NN$ interaction.
Thus one needs first an effective nonrelativistic parametrization
of GBE interaction which  would provide correct
nucleon mass, $N-\Delta$ splitting and the nucleon stability with this ansatz. 
For that one
can use the nonrelativistic parametrization \cite{GPP}
which satisfies approximately  the conditions above.

In order to have a qualitative insight into the  $NN$ interaction
it is convenient to use an adiabatic Born-Oppenheimer approximation
for the internucleon potential:

\begin{equation} V_{NN}(R) = <H>_R - <H>_\infty, \label{6} \end{equation}

\noindent
where $R$ is a collective coordinate which is the separation
distance between the two $s^3$ nucleons, $<H>_R$ is the lowest
expectation value of the 6Q Hamiltonian at fixed $R$, and $<H>_\infty$
is a mass of two well-separated nucleons ($2m_N$) calculated with the same
Hamiltonian.

At the moment we are interested in what is the $NN$ interaction at zero
separation between nucleons. It has been proved by Harvey that when
$R\rightarrow 0$, then in both
$^3S_1$ and $^1S_0$ partial waves in the $NN$ system only two types
of orbital 6Q configurations survive \cite{HARVEY}: 
$|s^6 [6]_O>$ and  $|s^4p^2 [42]_O>$, where $[f]_O$ is Young diagram,
describing spatial permutational symmetry in 6Q system. There are
a few different flavor-spin symmetries, compatible with the spatial symmetries
above: $[6]_O[33]_{FS}$, $[42]_O[33]_{FS}$, $[42]_O[51]_{FS}$, 
$[42]_O[411]_{FS}$, $[42]_O[321]_{FS}$, and $[42]_O[2211]_{FS}$. Thus,
in order to evaluate the $NN$ interaction at zero separation between
nucleons  it is necessary to
diagonalize a $6Q$ Hamiltonian in the basis above and use 
the procedure (\ref{6}).

From the adiabatic Born-Oppenheimer approximation (\ref{6}) we find
that $V_{NN}(R=0)$ is highly repulsive in both $^3S_1$ and $^1S_0$
partial waves, with the core being of order 1 GeV. 
This repulsion implies a strong suppression of the $NN$ wave function 
in the nucleon overlap region.

Due to the specific flavor-spin symmetry of GBE interaction between quarks 
the configuration
$s^4p^2[42]_O[51]_{FS}$ becomes highly dominant among other possible 6Q 
configurations at zero separation between nucleons (however, the "energy"
of this configuration is much higher than the energy of two well-separated
nucleons, that is why there is a strong short-range repulsion in 
$NN$ system). The symmetry structure of this
dominant configuration induces "an additional" effective repulsion,
related to the "Pauli forbidden state" in this case, and the s-wave 
$NN$ relative motion wave function has a node at short range\cite{NST}.
The existence of a strong repulsion, related to the energy ballance, discussed
above, suggests, however, that the amplitude of the
oscillating $NN$ wave function at short range will be strongly suppressed
as compared to that one of Moscow potential \cite{Mos}. 

Thus, within the chiral constituent quark model one has all the necessary
ingredients to understand microscopically the $NN$ interaction. There
appears strong short-range repulsion from the same short-range part of GBE
which also produces hyperfine splittings in baryon spectroscopy.
The long- and intermediate-range attraction in the $NN$ system is
automatically implied by the Yukawa part of pion-exchange and correlated
two-pion exchanges between quarks belonging to different nucleons.
With this first encouraging result, it might be worthwhile to perform
a more elaborate calculation of $NN$ and other baryon-baryon systems
within the present framework.

What will be a short-range interaction in other $YN$ and $YY$ systems?
In the chiral limit there is no difference between all octet baryons:
$N, \Lambda, \Sigma, \Xi$. Thus if one explains a strong short-range
repulsion in $NN$ system as related mostly to the spontaneous breaking
of chiral symmetry, as above, then the same short-range repulsion
should persist in other $YN$ and $YY$ systems. Of course, due to the
explicit chiral symmetry breaking the strength of this repulsion should
be essentially different as compared to that one in $NN$ system. One
can naively expect that it will be weaker.

\section{ "H-particle" should not exist}

Assuming that the
colour-magnetic interaction is a reason for hyperfine splittings
in baryons, Jaffe has predicted within the MIT bag model 
that there should be well bound dibaryon
in $\Lambda\Lambda$ system with $J^P=0^+$, called H-particle \cite{JAFFE}.
In this specific case the  colour-magnetic interaction in 6q system becomes
the most attractive and implies an existence of well-bound dibaryon, stable
against $\Lambda\Lambda$ strong decay, since it should be well below this
threshold. The H-particle should be a compact object in contrast to 
molecule-like state of nuclear nature - deuteron.

The quark-cluster calculations which rely on the constituent quark model
with a hybrid Hamiltonian (colour-magnetic interaction at short range and
meson-exchange at big distances) confirm this conclusion. With a simple
resonating group method variational ansatz they predict H-particle binding
energy of order of 20 or even 60-120 MeV \cite{RGM}. However, a simple
variational basis, used in these calculations, is rather poor at short
range (i.e. in 6Q region). As soon as it is properly extended, a very
deep bound state with binding energy of about 250 MeV has recently been
found \cite{Mal}.

Within our approach to baryon structure the situation is qualitatively
different. We have calculated the "energy" of what would be H-particle
and found it to be about 800 MeV above the $\Lambda \Lambda$ threshold
\cite{H}.
It then obviously suggests that there should not be H-dibaryon in Nature,
if our picture is adequate.
Thus the "H-particle" channel is not qualitatively different as compared
to other two-baryon systems.

This result is easy to anticipate from the symmetry properties. Only if
the colour-magnetic interaction is important for the hyperfine splittings,
like it is assumed in bag models or in traditional constituent quark model,
one could expect that the $[33]_{CS}$ colour-spin symmetry of the compact
6Q object (which is uniquely determined by the flavour-singlet nature of
"H-particle") is energetically more favourable than two well-separated
Lambdas. As soon as other physics, like in our case, is responsible for
hyperfine splittings in baryons, then the "H-particle" channel is not
energetically distinguished. So from this point of view a nonobservation
of "H-particle" in different experiments, extensively discussed at this
conference, gives an additional evidence against the colour-magnetic
nature of the hyperfine splittings in light and strange baryons.

\bigskip
\noindent
{\bf Acknowledgement}

It is my pleasure to thank D.O.Riska, Z.Papp, W.Plessas, K.Varga, 
R. Wagenbrunn, Fl. Stancu, and S. Pepin, in collaboration with whom 
different results
discussed in this talk have been obtained.

\end{document}